\documentclass[apj]{emulateapj}

\usepackage{lscape,graphicx,rotating}

\begin{document}

\title{GRB\,070125: The First Long-Duration Gamma-ray Burst in a Halo 
Environment\altaffilmark{1}}

\author{S.~Bradley Cenko\altaffilmark{2}, Derek B.~Fox\altaffilmark{3},
        Brian E.~Penprase\altaffilmark{4}, Antonio Cucchiara\altaffilmark{3},
	Paul A.~Price\altaffilmark{5}, Edo Berger\altaffilmark{6,7,8},
	Shri R.~Kulkarni\altaffilmark{9}, Fiona A.~Harrison\altaffilmark{2},
	Avishay Gal-Yam\altaffilmark{8},
	Eran O.~Ofek\altaffilmark{9}, Arne Rau\altaffilmark{9}, 
	Poonam Chandra\altaffilmark{10,11}, Dale A.~Frail\altaffilmark{12},
	Mansi M.~Kasliwal\altaffilmark{9}, 
	Brian P.~Schmidt\altaffilmark{13}, Alicia M.~Soderberg\altaffilmark{9},
	P.~Brian Cameron\altaffilmark{9}, Kathy C.~Roth\altaffilmark{14}}
\altaffiltext{1}{Based in part on observations obtained at the Gemini 
	Observatory which is operated by the Association of Universities for 
	Research in Astronomy, Inc., under a cooperative agreement with the 
	NSF on behalf of the Gemini partnership: the National Science 
	Foundation (US), the Particle Physics and Astronomy Research Council 
	(UK), the National Research Council (Canada), CONICYT (Chile), the 
	Australian Research Council (Australia), CNPq (Brazil) and CONICET 
	(Argentina).}
\altaffiltext{2}{Space Radiation Laboratory, MS 220-47,
        California Institute of Technology, Pasadena, CA 91125}
\altaffiltext{3}{Department of Astronomy and Astrophysics, Pennsylvania
        State University, 525 Davey Laboratory, University Park, PA 16802}
\altaffiltext{4}{Pomona College Department of Physics and Astronomy, 610
        N. College Ave, Claremont, CA 91711}
\altaffiltext{5}{Institute for Astronomy, University of Hawaii, 2680 Woodlawn
        Drive, Honolulu, HI 96822}
\altaffiltext{6}{Observatories of the Carnegie Institute of Washington,
        813 Santa Barbara Street, Pasadena, CA 91101}
\altaffiltext{7}{Princeton University Observatory, Peyton Hall, Ivy Lane,
        Princeton, NJ 08544}
\altaffiltext{8}{Astrophysics Group, Faculty of Physics, The Weizmann
	Institute of Science, Rehovot 76100, Israel}
\altaffiltext{9}{Division of Physics, Mathematics, and Astronomy, 105-24,
        California Institute of Technology, Pasadena, CA 91125}
\altaffiltext{10}{National Radio Astronomy Observatory, Charlottesville, VA,
	22903}
\altaffiltext{11}{Department of Astronomy, University of Virginia, P.O. Box
        3818, Charlottesville, VA 22903}
\altaffiltext{12}{National Radio Astronomy Observatory, Socorro, NM 87801}
\altaffiltext{13}{Research School of Astronomy and Astrophysics, Australian
        National University, Mt Stromlo Observatory, via Cotter Rd, Weston
	Creek, ACT 2611, Australia}
\altaffiltext{14}{Gemini Observatory, 670 N.~Aohoku Place, Hilo, HI 96720}

\email{cenko@srl.caltech.edu}

\slugcomment{\textit{ApJ} Accepted}

\shorttitle{The First GRB in a Halo Environment}
\shortauthors{Cenko \textit{et al.}}


\newcommand{\Swift}{\textit{Swift}}
\newcommand{\KW}{\textit{Konus-Wind}}
\newcommand{\HST}{\textit{Hubble Space Telescope}}
\newcommand{\gmos}{\textit{GMOS}}
\newcommand{\mgii}{\ion{Mg}{2} $\lambda \lambda$ 2796, 2803}
\newcommand{\mgi}{\ion{Mg}{1} $\lambda$ 2852}
\newcommand{\oiii}{[\ion{O}{3} $\lambda$ 3727]}

\begin{abstract}
We present the discovery and high signal-to-noise spectroscopic observations 
of the optical afterglow of the long-duration gamma-ray burst GRB\,070125.  
Unlike all previously observed long-duration afterglows in the redshift range 
$0.5 \lesssim z \lesssim 2.0$, we find no strong (rest-frame equivalent width 
$W_{\mathrm{r}} \gtrsim 1.0$\,\AA) absorption features in the wavelength range 
4000-10000\,\AA.  The sole significant feature is a weak doublet that 
we identify as \ion{Mg}{2} $\lambda \lambda$ 2796 ($W_{\mathrm{r}} = 
0.18 \pm 0.02$\,\AA), 2803 ($W_{\mathrm{r}} = 0.08 \pm 0.01$\,\AA) at 
$z = 1.5477 \pm 0.0001$.  The low observed \ion{Mg}{2} and inferred
\ion{H}{1} column densities are typically observed in galactic halos, 
far away from the bulk of massive star formation.  Deep ground-based 
imaging reveals no host directly underneath the afterglow to a limit of 
$R >$ 25.4 mag.  Either of the two nearest blue galaxies could host 
GRB\,070125; the large offset ($d \ge 27$\,kpc) would naturally explain 
the low column densities.  To remain consistent with the large local 
(i.e.~parsec scale) circum-burst density inferred from broadband 
afterglow observations, we speculate GRB\,070125 may have occurred far 
away from the disk of its host in a compact star-forming cluster.  Such 
distant stellar clusters, typically formed by dynamical galaxy interactions, 
have been observed in the nearby universe, and should be more prevalent at 
$z > 1$ where galaxy mergers occur more frequently.
\end{abstract}


\begin{deluxetable*}{llrrrrr}
  \tabletypesize{\footnotesize}
  \tablecaption{Log of Spectroscopic Observations}
  \tablecolumns{6}
  \tablewidth{0pc}
  \tablehead{\colhead{Identification} &
             \colhead{UT Date\tablenotemark{a}} &
             \colhead{Age\tablenotemark{b}} & \colhead{Primary Target} &
             \colhead{Wavelength Coverage} &
             \colhead{Airmass\tablenotemark{c}} &
             \colhead{Exposure Time} \\
             & & \colhead{(days)} & & \colhead{(\AA)} & &
             \colhead{(s)}
            }
  \startdata
  1a & 2007 Jan 26.228 & 0.922 & Afterglow & 5900-10000 & 1.98 & 1800 \\
  1b & 2007 Jan 26.250 & 0.944 & Afterglow & 5900-10000 & 1.67 & 1800 \\
  1c & 2007 Jan 26.272 & 0.966 & Afterglow & 4000-8100 & 1.45 & 1800 \\
  1d & 2007 Jan 26.294 & 0.989 & Afterglow & 4000-8100 & 1.30 & 1800 \\
  2a & 2007 Jan 29.247 & 3.941 & Afterglow & 5900-10000 & 1.60 & 1800 \\
  2b & 2007 Jan 29.269 & 3.963 & Afterglow & 5900-10000 & 1.41 & 1800 \\
  2c & 2007 Jan 29.290 & 3.984 & Afterglow & 5900-10000 & 1.27 & 1800 \\
  2d & 2007 Jan 29.312 & 4.006 & Afterglow & 5900-10000 & 1.18 & 1800 \\
  3a & 2007 Feb 4.340 & 9.992 & $R1$ & 5900-10000 & 1.06 & 2400 \\
  3b & 2007 Feb 4.369 & 10.063 & $R1$ & 5900-10000 & 1.03 & 2400 \\
  \enddata
  \tablenotetext{a}{UT at midpoint of exposure.}
  \tablenotetext{b}{Age in days from detection of the burst at 7:20:45 UT on
        2007 January 25 \citep{GCN.6024}.}
  \tablenotetext{c}{Average airmass of exposure.}
\label{tab:spec}
\end{deluxetable*}

\section{Introduction}
\label{sec:intro}

The connection between long-duration ($\Delta t \gtrsim 2$ s) gamma-ray bursts
(GRBs) and hydrogen-stripped, core-collapse supernovae (i.e.~Type Ib/c SNe) 
is now well-established in the nearby universe (see e.g.~\citealt{wb06}).  
At $z \gtrsim 0.3$, where the overwhelming majority of GRBs are detected
(e.g.~\citealt{bkf+05,jlf+06}), Type Ib/c SNe are too faint, 
absorbed, and redshifted to be observed routinely with current facilities.  
Observations of the \textit{environments} of distant GRBs, however, are
consistent with a massive star origin.  GRB hosts are typically 
faint, blue, irregular galaxies with large specific 
star-formation rates (star formation rate per unit stellar mass; 
\citealt{ldm+03,chg04}).  And within their hosts, GRB afterglows are found to
be concentrated in the innermost regions, tracing the blue light
from hot young stars even more strongly than Type Ib/c SNe 
\citep{bkd02,fls+06}.

Bright GRB afterglows are therefore ideally suited to probe the dense gas
in the very regions where stars are being formed.  This stands in marked
contrast to quasar (QSO) sight lines, which sample galaxies according to gas
cross-section and are therefore much more likely to probe the outer regions
of galaxy halos (e.g.~\citealt{pcd+07}).

While the sample of afterglow absorption spectra suitable for
elemental abundance studies is still quite small compared with QSOs, a
general picture has nonetheless begun to take hold.  
GRB systems are characterized by:
1) large metal equivalent widths and correspondingly large metal
column densities (e.g.~\citealt{mdk+97}), 2) extremely high neutral hydrogen 
column densities (e.g.~\citealt{hmg+03}), typically falling at log 
$N$(\ion{H}{1}) $> 20.3$ (the so-called damped Lyman-$\alpha$ systems, 
or DLAs; \citealt{wgp05}), and 3) sub-solar metallicities, typically 
$Z \sim 0.1\,Z_{\odot}$ (e.g.~\citealt{bpc+06}).  All three findings are 
consistent with a massive star origin for long-duration GRBs.

Here we present observations of a long-duration event, GRB\,070125, 
that does not fit neatly into this paradigm.  Despite deep 
spectroscopy of a bright ($R \approx 19$ mag) afterglow, we detect only
weak \ion{Mg}{2} absorption at $z = 1.55$, a firm upper limit on the 
\ion{Mg}{2} column density of the host galaxy.  Coupled with the large offset 
between the afterglow and the nearest detected host galaxy candidate, our 
observations indicate the large-scale (i.e.~ISM) burst environment 
is dramatically different from all previously observed GRB hosts.

\section{Observations and Data Reduction}
\label{sec:obs}

GRB\,070125 was discovered by the Inter-Planetary Network 
at 07:20:45 UT on 2007 January 25 \citep{GCN.6024}.  The burst 
was notable both for its brightness (fluence $= 1.75^{+0.18}_{-0.15}
\times 10^{-4}$\,erg\,cm$^{-2}$; \citealt{GCN.6049}) and its long duration 
($\Delta t \gtrsim 200$\,s; \citealt{GCN.6024,GCN.6049}).  The
well-characterized prompt emission allowed measurements of the peak energy
of the spectrum ($E_{\mathrm{peak}} = 367^{+65}_{-51}$\,keV; 
\citealt{GCN.6049}), as well as a ``pseudo-redshift'' \citep{ph06} of
$z_{\mathrm{p}} = 1.3 \pm 0.3$ \citep{GCN.6059}.

\begin{figure*}[t]
        \centerline{\plotone{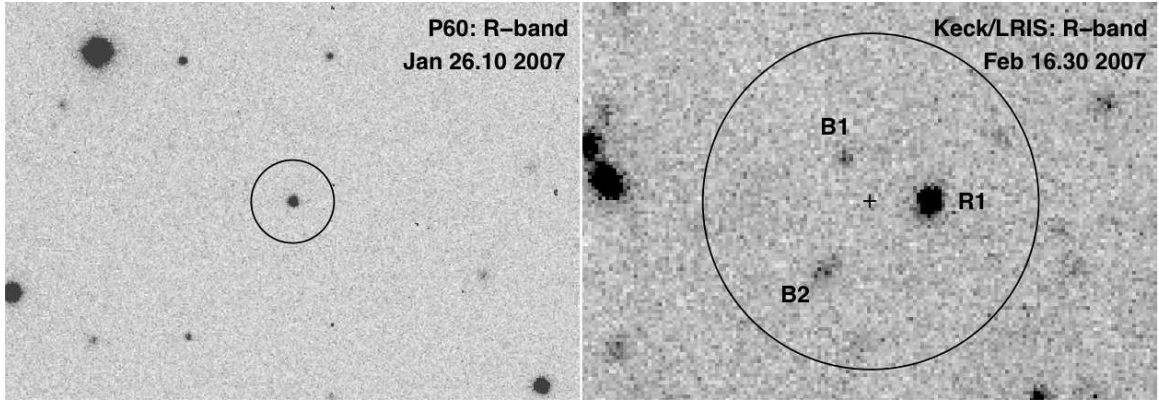}}
        \caption[Optical imaging of the field of GRB\,070125]
        {Optical imaging of the field of GRB\,070125.  \\
        \textit{Left}: P60 $R$-band discovery image of the optical afterglow of
        GRB\,070125.  The afterglow is centered inside a circle of radius
        10\arcsec.  \\
        \textit{Right}: Late-time Keck/LRIS $R$-band image
        of the field.  The location of the afterglow is marked with a cross.
        The black circle again has a radius of 10\arcsec.  We find no
        evidence for an underlying host, to limits of $R > 25.4$ mag,
        $g^{\prime} > 26.1$ mag.  The three galaxies nearest to the
        afterglow location are marked ($B1$, $B2$, and $R1$).  $R1$,
        initially suggested as a possible host for GRB\,070125
        \citep{GCN.6054}, is a red, foreground galaxy ($z = 0.897$) with little
        current star formation.  $B1$ and $B2$ are both quite blue
        ($g^{\prime} - R \approx 0$), typical of long-duration GRB hosts.
        All images are oriented with north up and east to the left.}
\label{fig:finder}
\end{figure*}

We began observing the field of GRB\,070125 with the automated Palomar 
60-inch telescope \citep{cfm+06} at 02:18:59 UT on 2007 January 26 ($\Delta 
t = 19.0$ hours).  Inside the burst error circle, we found a bright, stationary 
source ($R = 18.59 \pm 0.03$) not present in the Sloan Digital Sky 
Survey images of this field \citep{aaa+06} that we identified as the 
optical afterglow of GRB\,070125 (\citealt{GCN.6028}; Fig.~\ref{fig:finder}).  
Our subsequent broadband monitoring of the afterglow of GRB\,070125 is 
presented in a separate work (Chandra et al.~2007, in preparation).

We also undertook spectroscopic observations of GRB\,070125 with the Gemini 
Multi-Object Spectrograph (GMOS; \citealt{hab+03}) mounted on the 8-m Gemini 
North Telescope beginning on the night of 2007 January 26.  
For all spectra, we employed a $2\times2$ binning to increase
the CCD signal-to-noise ratio, and we used the R400 grating and
1\arcsec\ slit.  Our configuration resulted in a spectral resolution of
$\sim 8$\,\AA\ and a dispersion of $1.34$\,\AA\,pix$^{-1}$.
The details of our observations are shown in Table \ref{tab:spec}.  

All spectra were reduced in the IRAF\footnote{IRAF is distributed by the
National Optical Astronomy Observatory, which is operated by the 
Association for Research in Astronomy, Inc., under cooperative agreement with
the National Science Foundation.} environment using standard
routines.  Pairs of dithered spectra were subtracted to remove residual 
sky lines.  Cosmic rays were removed using the LA Cosmic routine 
\citep{v01}.  Spectra were extracted optimally \citep{h86} and wavelength 
calibration was performed first relative to CuAr lamps and then tweaked based 
on night sky lines in each individual image.  In all cases, the resulting RMS 
wavelength uncertainty was $\lesssim 0.3$\,\AA.  Both air-to-vacuum and 
heliocentric corrections were then applied to all spectra.  Extracted spectra 
were divided through by a smoothed flux standard to remove narrow band 
($< 50$\,\AA) instrumental effects \citep{b99}.  Finally, telluric atmospheric 
absorption features were removed using the continuum from spectrophotometric 
standards \citep{wh88,mfh+00}.

Deep, late-time imaging to search for the host galaxy of GRB\,070125 was 
taken with the Low-Resolution Imaging Spectrometer (LRIS: \citealt{occ+95})
mounted on the 10-m Keck I telescope.  LRIS employs a dichroic, allowing 
simultaneous imaging in both $g^{\prime}$- and $R$-band filters.
We obtained 4 $\times$ 300 s images at a mean epoch of 7:12:06.6 UT on 2007 
February 16.  Individual images were bias-subtracted and flat-fielded using
standard IRAF routines.  Co-addition was performed using 
SWarp\footnote{http://terapix.iap.fr}.  The resulting $R$-band 
image is shown in Figure 1 (right panel). 

\section{Results}
\label{sec:results}
\begin{figure*}[t]
        \centerline{\plotone{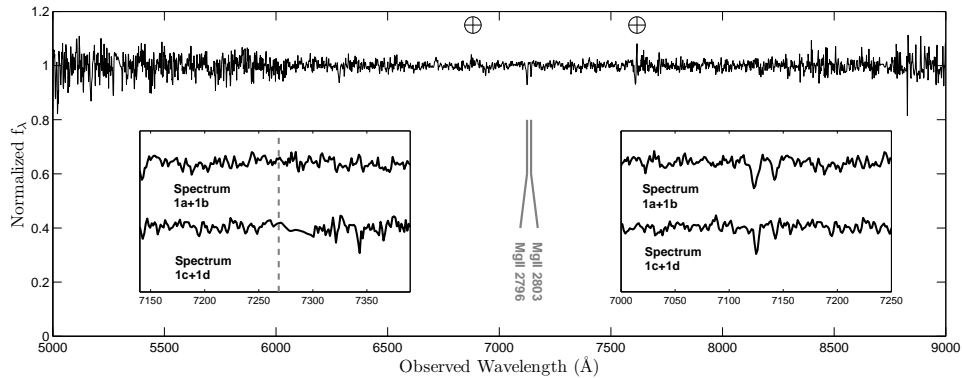}}
        \caption[GMOS spectrum of GRB\,070125]
        {Gemini/GMOS spectrum of the afterglow of GRB\,070125.  In the
        primary plot we show the sum of all four spectra obtained the night
        of 2007 January 26 after normalizing by the continuum (1a-1d; see
        Table \ref{tab:spec}).  The region
        from 6000-8000 \AA\ shows the highest signal-to-noise ratio, as it was
        covered in all four spectra.  Telluric atmospheric absorption features
        are indicated by circled plus signs.  No strong absorption features
        are present in the spectrum.  The strongest feature is a doublet
        at 7124/7142 \AA, which we identify as \mgii\ at $z = 1.55$.  Despite
        its weakness, the doublet is visible in spectra from both
        configurations taken on 2007 January 26 (right inset).  In the left
        inset, we zoom in on the \mgi\ transition at $z = 1.55$.
        No absorption is detected in either configuration to $W_{\mathrm{r}}
        < 0.06$\,\AA.}
\label{fig:spec}
\end{figure*}

In Figure \ref{fig:spec} we show a sum of all four GMOS spectra 
obtained on the night of 2007 January 26 (1a-1d; see Table \ref{tab:spec}),
normalized by the continuum.  The strongest absorption feature present is a 
doublet at $\lambda \lambda$ 7124,7142\,\AA, with observed equivalent widths of 
$0.47 \pm 0.05$\,\AA\ ($\lambda_{1} = 7124.23 \pm 0.35$\,\AA) and 
$0.21 \pm 0.04$\,\AA\ ($\lambda_{2} = 7142.45 \pm 0.42$\,\AA).  
Despite the weakness of the feature, the doublet 
is detected in separate co-additions of spectra from the two different 
instrumental configurations ($\lambda_{\mathrm{central}} = 6000$\,\AA\
vs.~$\lambda_{\mathrm{central}} = 8000$\,\AA), providing strong
confirmation of its reality (see Fig.~\ref{fig:spec}, right inset).  
Inspection of the two-dimensional spectra reveals the only other significant
absorption feature, at $\lambda = 6283$\,\AA, is offset slightly from the 
center of the trace.  Furthermore, it is only detected in one instrumental
configuration.  We therefore believe 
this feature is most likely an artifact of the data reduction process.

We find marginal ($\lesssim 2\sigma$) evidence for strengthening of both 
features in the doublet over the duration of our observations.  
Variability has been reported before in GRB afterglow spectra, both within 
the GRB host itself (time-dependent excitation caused by UV photons from the
GRB afterglow; \citealt{vls+07}) and in intervening absorbers (caused by
variability in the GRB beam size relative to intervening clouds; 
\citealt{hsd+07}).  However, because of the uncertainty of this result, we
proceed using average values measured from the sum of all our spectra
obtained on the night of 2007 January 26.

Based on the lack of Lyman-$\alpha$ absorption, we place an upper limit
on the afterglow redshift of $z \le 2.3$.  \citet{GCN.6032} report a 
contemporaneous LRIS spectrum of the afterglow with coverage extending down
to the atmospheric cut-off at $\lambda \approx 3000$\,\AA.  Based on the 
absence of damped Lyman-$\alpha$ absorption or Lyman-$\alpha$ forest emission,
they report a redshift upper limit of $z < 1.4$.  Given the weakness of the
observed doublet, the expected weakness of the associated 
Lyman-$\alpha$ absorption (see below), and the decrease in sensitivity at 
the bluest LRIS wavelengths, we believe this limit is too strict.  Instead,
we adopt a more conservative value of $z < 1.8$ (corresponding to
$\lambda_{\mathrm{Ly-}\alpha} \le 3500$\,\AA) throughout this work.

Consistent with the above redshift constraint, we identify the observed
doublet as \mgii\ at $z = 1.5477 \pm 0.0001$.  Besides the observed
wavelength ratio, we offer two additional pieces of evidence in support
of this association.  First, the observed equivalent width ratio
(\ion{Mg}{2} $\lambda$ 2796 / \ion{Mg}{2} $\lambda$ 2803) is consistent with 
the value of 2:1 predicted for weak, unsaturated absorption from this 
transition.  Second, the \mgii\ doublet is the strongest absorption feature 
observed in \textit{all} GRB hosts identified in the redshift range 
$0.5 \lesssim z \lesssim 2.0$ (see below).  
Even if the system does not arise from the GRB host, this doublet is commonly 
found in intervening systems of both QSOs \citep{ss92} and GRBs 
\citep{ppc+06}.  At this redshift, we place an upper
limit on the rest-frame equivalent width of absorption from \mgi\ of
$W_{\mathrm{r}} < 0.06$\,\AA\ (Fig.~\ref{fig:spec}, left inset).

Based on the observed \ion{Mg}{2} equivalent widths, we can calculate 
corresponding column densities in the optically thin (i.e.~unsaturated) limit:
\begin{equation}
N = \frac{m_{e} c^{2}}{\pi e^{2}} \frac{W_{\mathrm{r}}}{f \lambda^{2}} = 
	1.13 \times 10^{20} \mathrm{cm}^{-2} 
	\frac{(W_{\mathrm{r}}/\mathrm{\AA})} {(\lambda / \mathrm{\AA})^{2} f},
\label{eqn:optthin}
\end{equation}
where $f$ is the oscillator strength (from \citealt{m91}),
$W_{\mathrm{r}}$ is the rest-frame equivalent width, and $\lambda$ is the rest
wavelength.  The results are shown in Table~\ref{tab:lineids}.  For the 
observed system we measure a column density of log $N$(\ion{Mg}{2}) 
$= 12.61 \pm 0.05$.  The corresponding upper limit on the \ion{Mg}{1} column 
density is log $N$(\ion{Mg}{1}) $< 11.7$.

To compare our observed \ion{Mg}{2} system with previous samples, we would 
like to know whether it arises from the GRB host or some intervening galaxy.  
To this end, we undertook a second epoch of GMOS spectroscopy on the night of 
2007 January 29 (2a-2d; see Table \ref{tab:spec}) to search for nebular 
emission lines \textit{at the location of the afterglow}.  While still bright 
enough to provide a reliable trace ($R \sim 21.5$) at this epoch, the decreased 
afterglow flux improved our sensitivity to faint emission lines.  
At $z = 1.5477$, the only common line indicative of active star formation to 
fall in our bandpass is \oiii\ ($\lambda_{\mathrm{obs}} \approx 9497$).  
The presence of several bright night sky lines nearby significantly affected 
our sensitivity.  Nonetheless, we put an upper limit on the observed 
flux\footnote{Throughout this work, we adopt a $\Lambda$CDM cosmology
with the latest parameters from \textit{WMAP} ($H_{0} = 70.9$ km s$^{-1}$
Mpc$^{-1}$; $\Omega_{\mathrm{m}} = 0.266$; $\Omega_{\Lambda} = 1 -
\Omega_{\mathrm{m}}$; \citealt{sbd+07}).} from
\oiii\ at $z = 1.5477$ of $< 5 \times 10^{-18}$ erg cm$^{-2}$ s$^{-1}$.  
Using the relation from \citet{k98}, this corresponds to an upper limit on the
star-formation rate of $< 1$ M$_{\odot}$ yr$^{-1}$.  This lies on the low
end of star-formation rates observed in previous GRB hosts \citep{chg04}.

Without a secure emission-line redshift, we cannot determine the nature
of the \ion{Mg}{2} system (i.e.~host or intervening).  Nonetheless, because of
our redshift constraints, $1.55 \le z \le 1.8$, the \mgii\
transition from the host
is guaranteed to fall within our observed bandpass (n.b.~this is the case
even if we apply our weaker $z \le 2.3$ constraint).  Therefore,
\textit{even if the observed \ion{Mg}{2} system is from an intervening galaxy,
the measured equivalent widths are a firm upper limit on the presence of
\ion{Mg}{2} in the host system}.  In what follows we shall assume that
$z = 1.55$ is the redshift of the GRB host; all our conclusions below are
only strengthened if $z > 1.55$.

\begin{figure*}[t]
        \centerline{\plotone{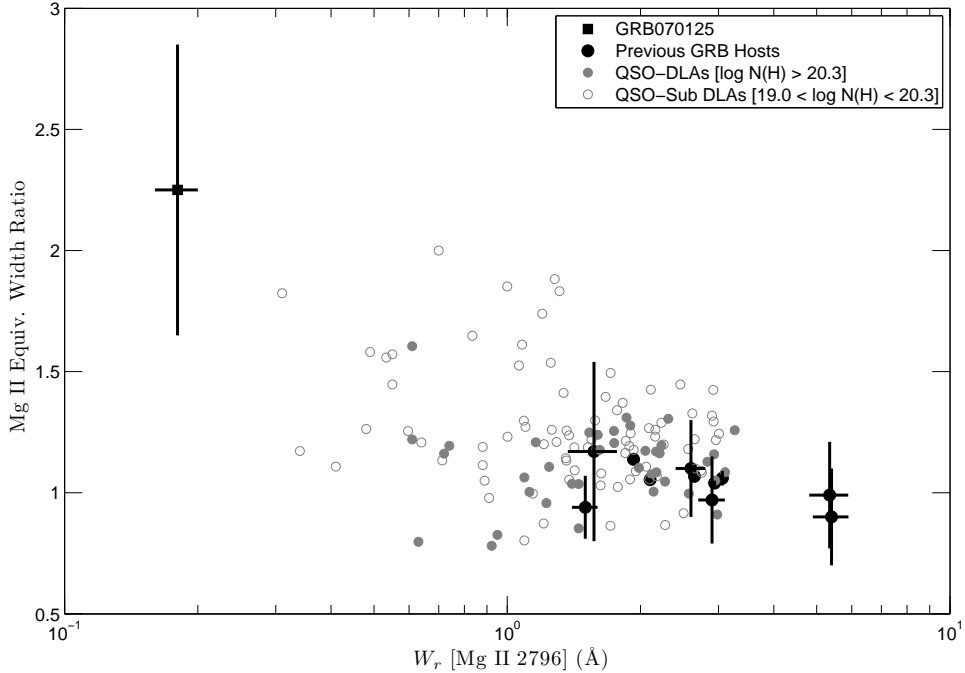}}
        \caption{\ion{Mg}{2} absorption in GRB host galaxies.  Here we
        plot a compilation of all the equivalent width measurements of the
        \ion{Mg}{2} $\lambda$ 2796 absorption feature in GRB host galaxies.
        On the y-axis, we plot
        the observed ratio between the \ion{Mg}{2} $\lambda$ 2796 and
        \ion{Mg}{2} $\lambda$ 2803 absorption
        lines.  Ratios deviating from 2 indicate the lines have become
        saturated and the corresponding optically thin column densities should
        be treated as lower limits.  Thus the factor of $\approx 10-15$
        discrepancy between GRB\,070125 and all previous GRB hosts actually
        underestimates the true difference in column densities.  Shown
        in gray are analogous measurements for QSO-DLAs (log $N$(\ion{H}{1})
        $> 20.3$; filled circles) and QSO-Sub DLAs ($19.0 <$ log
        $N$(\ion{H}{1}) $< 20.3$; empty circles).  References:
        GRB\,970508: \citet{mdk+97}; GRB\,990123: \citet{kdo+99};
        GRB\,000926: \citet{cgh+03}; GRB\,010222: \citet{mhk+02};
        GRB\,020813: \citet{bsc+03}; GRB\,030226: \citet{sbp+06,kgr+04};
        GRB\,030328: \citet{mmp+06}; GRB\,041006: \citet{skp+06};
        GRB\,051111: \citet{pbf+06,pcb+07}; GRB\,060418: \citet{vls+07,pcb+07};
        GRB\,070208: Cucchiara et al.~(2007, in preparation);
        QSOs: \citet{rtn06}.}
\label{fig:mgII}
\end{figure*}

In Figure \ref{fig:mgII} we plot
the observed \ion{Mg}{2} $\lambda$ 2796 rest-frame equivalent width 
for GRB\,070125 compared to all 
previously observed GRB hosts. 
On the ordinate axis we plot the ratio between the 2796\,\AA\ 
and 2803\,\AA\ components of the doublet.  We find GRB\,070125 is an outlier 
on both axes.  Clearly the inferred column density is significantly lower 
than any other GRB host galaxy.  Furthermore, the \ion{Mg}{2} $\lambda$ 2796 /
\ion{Mg}{2} $\lambda$ 2803 ratio in all previous GRBs was $\approx 1$, 
indicating significant saturation.  For saturated lines, the optically thin 
approximation (eq.~\ref{eqn:optthin}) breaks down and can significantly 
underestimate the true column density.  The difference in \ion{Mg}{2} 
column density between the host galaxy of GRB\,070125 and all previous 
GRB hosts is therefore even larger than the factor of $\approx 10-15$ 
derived above.   

To verify that the sample of GRB hosts used above is not biased toward
strong \ion{Mg}{2} absorbers, we searched through the GRB Circulars Network
(GCN\footnote{http://gcn.gsfc.nasa.gov/gcn3\_archive.html.}) archive to 
review all reported long-duration afterglow absorption spectra.  Neglecting
the most nearby events, $z \lesssim 0.3$, for which the strongest absorbers 
still lie in the UV, we find only a single report of a high signal-to-noise
spectrum absent any absorption features in the optical bandpass
(GRB\,061021; \citealt{GCN.5747}).  Alternatively, of the 17 long-duration
events with a reported redshift and
spectral coverage of the host \mgii\ transition, but without
reported equivalent widths (i.e.~not included in Figure \ref{fig:mgII}),
16 report a 
detection of this doublet.  The sole exception, a spectrum of GRB\,050802, 
contains several absorption features, but their identification was 
uncertain \citep{GCN.3749}.  We can therefore rule out a significant 
population of weak \ion{Mg}{2} absorbers from GRB hosts at a high degree of 
confidence.

The detection of \ion{Mg}{1} from GRB hosts is thought to indicate
that these observations probe distances far away ($\ge 100$\,pc)
from the GRB itself \citep{pcd+07,vls+07}.  The first ionization energy 
of Mg is 7.6\,eV, and therefore UV photons from the GRB afterglow are
able to ionize any \ion{Mg}{1} in the circum-burst medium to \ion{Mg}{2}
(subsequent ionizations beyond \ion{Mg}{2} are likely shielded
by neutral H, as their ionization energies lie above 1\,Ryd).  It is
therefore important to determine if we would expect to see \ion{Mg}{1}
absorption from GRB\,070125, or whether the feature would be too weak
to detect in our spectra.

Comparing the equivalent width ratio of \ion{Mg}{2} $\lambda$ 2803
(the weaker of the \ion{Mg}{2} doublet, and therefore less saturated)
to that of \ion{Mg}{1} $\lambda$ 2853 in previously observed GRB hosts,
we find ratios ranging from 1.7 (GRB\,970508; \citealt{mdk+97}) to
2.8 (GRB\,060418; \citealt{pcb+07,vls+07}).  Again we note these values
are really a lower limit, as saturation will be more significant for the
stronger \ion{Mg}{2} $\lambda$ 2803 feature.  Nonetheless, we predict 
an \ion{Mg}{1} $\lambda$ 2853 rest frame equivalent width of $\lesssim
0.03 - 0.05$\,\AA.  This is below our sensitivity limit, and we therefore
believe we would not be sensitive to \ion{Mg}{1} absorption even if it
were present at expected levels.

To convert our measured Mg column density (assumed to be dominated by 
\ion{Mg}{2}) to neutral hydrogen, we must estimate the ratio of Mg:H in the 
host.  Previous GRB hosts range in metallicity from $-2.0 \lesssim 
[\mathrm{M/H}] \lesssim -0.5$ \citep{bpc+05,pcd+07}, much like QSO-DLAs 
\citep{pgw+03}.  At $z < 2$, no QSO-DLA has ever been observed with [M/H] 
$< -2$ \citep{wgp05}, and a near-solar metallicity is difficult to reconcile
with a star-forming galaxy at $z=1.55$.  If we neglect the cold, dense, 
disk depletion model, ruled out for all GRB hosts observed to date 
\citep{sff03,bpc+05,pbf+06}, all other environments predict an Mg depletion 
of [Mg/M] $\approx -0.5$ \citep{ss96}.  With the above limits, we estimate
the neutral hydrogen column density to fall within $18.0 \lesssim$ 
log $N$(\ion{H}{1}) $\lesssim 19.5$ (using solar abundances from 
\citealt{ags05}).

The implied Mg:H ratio is in good agreement with previously observed
GRBs, as well as the broader sample of QSOs.  While $N$(\ion{Mg}{2}) and 
$N$(\ion{H}{1}) have never been accurately measured simultaneously in a 
single GRB host, two events provide upper limits: [\ion{Mg}{2}/\ion{H}{1}] 
$> -2.7$ for GRB\,000926 \citep{fgm+02,cgh+03}, and [\ion{Mg}{2}/\ion{H}{1}] 
$> -1.6$ for GRB\,030226 \citep{kgr+04,sbp+06}.  For QSO-DLAs at $z < 1.65$, 
the mean equivalent width for the \ion{Mg}{2} $\lambda$ 2796 transition is 
$1.8$\,\AA, while for sub-DLAs ($19.0 \le$ log $N$(\ion{H}{1}) $\le 20.3$) 
the corresponding value is $<W_{\mathrm{r}}> = 1.6$\,\AA\ \citep{rtn06}.  
For comparison, we also plot these QSO-\ion{Mg}{2} systems in Figure 
\ref{fig:mgII}.  In a sample of eight weak ($W_{\mathrm{r}} < 0.3$\,\AA) 
\ion{Mg}{2} absorbers, \citet{crc+99} found all had log $N$(\ion{H}{1}) 
$< 19.0$.  In fact, six of the eight systems exhibited no sign of a Lyman 
limit break, indicating not only log $N$(\ion{H}{1}) $< 17$, but also that 
neutral hydrogen was optically thin in those clouds \citep{t82}.  

An alternative way to determine the neutral hydrogen density is by modeling
the afterglow spectral energy distribution (SED).  Our models
of the X-ray spectrum and optical SED do not require any dust extinction
in addition to the Galactic component (Chandra et al. 2007, in preparation; see
also \citealt{GCN.6030}).  However,
because of the relatively large Galactic column ($N$(H) $\approx 5 \times
10^{20}$\,cm$^{-2}$; \citealt{dl90}) and the large effect of redshift on dust
obscuration, these limits are not nearly 
as constraining as those derived from the optical spectrum.  

All told, we have strong evidence that the neutral hydrogen column density
in the host of GRB\,070125 is quite low: log $N$(\ion{H}{1}) $< 19.5$, if not
significantly smaller.  Such densities are usually associated with galaxy
halos, and stand in contrast with the sample of previously observed GRB hosts,
which are commonly attributed to a disk population.  The sample of previously 
observed GRB hosts has a median log $N$(\ion{H}{1}) $\approx 21.3$, with an 
observed standard deviation of 0.9 dex \citep{jfl+06}.  In other words, to fall 
within one sigma of the known distribution, a GRB host must be a DLA.  Only
three previous events, GRBs 021004 \citep{mfh+02}, 050908 \citep{GCN.3949},
and 060607 \citep{jfl+06} had measured log $N$(\ion{H}{1}) $< 19.5$, and 
none report significantly lower values.  All three exhibited strong absorption 
from other metals in addition to Lyman-$\alpha$, again distinguishing them
from the mostly featureless spectrum of GRB\,070125.  

In our late-time imaging of the field of GRB\,070125 (Fig.~\ref{fig:finder},
right panel), we find no host directly underneath the afterglow location
to limits of $R > 25.4$ mag (Vega), $g^{\prime} > 26.1$ mag (AB).  Using
a synthetic spectrum of a star-forming galaxy from \citet{kcb+96}, we
estimate a limit on the absolute magnitude of $M_{\mathrm{V}} > -19.2$ mag
for any underlying host.  Many GRB host galaxies are fainter than
$M_{\mathrm{V}} > -19.2$ \citep{fls+06}, so it is entirely possible our 
limits are too shallow to detect the underlying emission.  Nonetheless, 
because of the low density environment, we also consider the
possibility that the afterglow lies significantly further away from its
host than the typical GRB ($\le 10$ kpc; \citealt{bkd02,fls+06}).

We identify three candidate host galaxies within 10\arcsec\ of the afterglow
location: $R1$, $B1$, and $B2$ (see Fig.~\ref{fig:finder}, right panel).
$R1$, 3.6\arcsec\ to the west of the afterglow, is a large red 
($g^{\prime} - R \approx 2.4$ mag) galaxy identified by \citet{GCN.6054} 
as a possible host for GRB\,070125.  Based on a Gemini-GMOS spectrum obtained
on the night of 2007 February 4 (3a-3b; see Table \ref{tab:spec}), 
we identify a strong continuum break
at $\sim 7500$\,\AA\ as the rest frame 4000 \AA\ break.  Ca H+K and G-band
absorption confirm the galaxy lies in the foreground at 
$z=0.897$.\footnote{At a distance of 3.6\arcsec, $R1$ would need to be
extremely massive ($\sim 10^{13}$ M$_{\odot}$) for gravitational lensing to
significantly affect the afterglow.}  The other two objects, $B1$ at 
3.2\arcsec\ distance, and $B2$ at a distance of 5.5\arcsec, are both blue
($g^{\prime} - R \approx 0$ mag) and compact, more typical of long-duration
GRB hosts \citep{ldm+03}.  Unfortunately neither galaxy fell on the slit in 
any of our spectra.

\section{Discussion}
\label{sec:disc}

Finally, we speculate on the origin of GRB\,070125.  At first glance, a 
compact binary progenitor system, as has been argued to explain most 
short-duration bursts \citep{elp+89} seems appealing for GRB\,070125: the 
large host offset and low density environment could naturally be explained 
by the asymmetric ``kick'' imparted to such systems as the members become 
SNe \citep{fwh99,bsp99}.  The accretion disk formed in such a merger is
expected to last only a fraction of a second \citep{npk01}; a serious 
discrepancy with the observed duration.  However, the recent discovery of two 
nearby, long-duration GRBs lacking associated SN emission
\citep{gfp+06,fwt+06,dcp+06,ocg+07} leads us to at least consider an origin
not associated with massive stars.

In a separate work, Chandra et al.~(2007, in preparation) 
study the broadband afterglow emission
from GRB\,070125.  Two findings from this study cast doubt on a compact
binary merger origin for this event.  First, the total energy
release from GRB\,070125, including the collimation correction, is
extreme even for long bursts ($E \gtrsim 10^{52}$ erg).
Short-duration bursts typically are less energetic ($E \lesssim 10^{50}$ erg;
\citealt{ffp+05}), although the higher-redshift examples discussed in
\citet{bfp+07} appear to be more luminous, and may call this into question.

More importantly, however, based on the broadband SED 
(particularly the self-absorbed radio spectrum), we conclude the
\textit{local} (parsec scale) circum-burst density is quite high,
even for typical long-duration afterglows ($n \sim 20$ cm$^{-3}$ for
a constant density environment).
While this may seem inconsistent with the low \ion{Mg}{2} column
density derived from absorption spectroscopy, we instead consider the two
observations the strongest evidence to date that afterglow studies
and absorption spectroscopy probe \textit{distinct regions}: the parsec-scale
circum-burst medium for the afterglow vs.~the more distant ($\ge 100$ pc)
ISM for absorption spectroscopy \citep{pcd+07,vls+07}.  
In the compact binary merger
scenario, a large host offset should imply a low circum-burst 
density ($n \lesssim 10^{-3}$ cm$^{-3}$), as has been seen for many 
short-duration bursts already \citep{ffp+05,sbk+06,p06}.  

Because of the long duration, large local density, and large energy release,
we return again to consider a massive star progenitor.  Instead, we now
must explain how a massive star could end up so far away from the dense disk
of its host.  For the closest putative host from our LRIS imaging, the observed
offset of 3.2\arcsec\ corresponds to a projected distance of $\approx 27$ kpc 
at $z = 1.5477$.  To travel this distance in its short lifetime, a massive star
would need an extremely large peculiar velocity: $\sim 10^{4}$ km s$^{-1}$ for 
a 20 Myr lifetime.  The fastest known object in the galaxy is the
Guitar pulsar, with a peculiar velocity of 1600 km s$^{-1}$ \citep{crl93}, 
while galactic stars have been identified with peculiar velocities as large
as 500 km s$^{-1}$ (presumably accelerated by interacting with a black hole;
\citealt{bgk+06}).  It is much more probable that the progenitor was formed 
\textit{in situ}.

Such a scenario has precedent in the local universe, where young, massive,
compact star clusters have been found at large distances (i.e.~several times
the optical radius) either in extended UV disks 
(e.g.~M83: \citealt{tbb+05}, NGC4625: \citealt{gmb+05}) or in tidal tails
of interacting galaxies (e.g.~``Antenna'' system: \citealt{hbt+05,zfw01},
``Tadpole'' galaxy: \citealt{jpf+06}).  In some of the most strongly
interacting systems, $\ge 10$\% of the current star formation is occurring
in such clusters \citep{jpf+06}.  
Broadband surveys of nearby galaxies indicate a
significant fraction ($\lesssim 1$\%) of the current star formation in the
local universe takes place in these extreme environments (D.~Calzetti, private
communication).
With our current understanding of hierarchical galaxy formation, such
interactions should only increase in frequency as a function of look-back time.
In retrospect, it is not entirely surprising that, of the $\sim 50$
long-duration GRBs with absorption spectra, we should discover such an
event.  While a thorough discussion of the relative frequency of such events
is premature, the rarity of events like GRB\,070125 implies star formation
in the outer regions of galaxies in the distant universe is likely not 
dramatically different from what we observe today.

\acknowledgments
We wist to thank R.~Ellis for obtaining the Keck ToO imaging data.
Some of the data were obtained with the Gemini Observatory under
Program IDs GN-2006B-Q-21 and GN-2007A-Q-3.
Some of the data presented herein were obtained at the
W.~M.~Keck Observatory, which is operated as a scientific partnership
among the California Institute of Technology, the University of
California and the National Aeronautics and Space Administration.
The Observatory was made possible by the generous financial support of
the W.~M.~Keck Foundation.
S.~B.~C.~and A.~M.~S.~are supported by the NASA Graduate
Student Research Program.  E.~B.~is supported by NASA through Hubble
Fellowship grant HST-HF-01171.01 awarded by STScI, which is operated
by the Association of Universities for Research in Astronomy, Inc., for
NASA, under contract NAS5-26555.  A.~G.~acknowledges support by NASA through
Hubble Fellowship grant HST-HF-01158.01 awarded by STScI.  P.~C.~is
supported by a Jansky fellowship.  M.~M.~K.~is supported by the 
Gordon and Betty Moore Foundation via a Hale Fellowship.  GRB research
at Caltech is supported by NASA and the NSF.  The authors wish to recognize and 
acknowledge the very significant cultural role and reverence that the summit 
of Mauna Kea has always had within the indigenous Hawaiian community.  
We are most fortunate to have the opportunity to conduct observations from 
this mountain.  

{\it Facilities:} \facility{Gemini:Gillett (GMOS)}, \facility{Keck:I (LRIS)}, 
\facility{PO:1.5m ()}

\bibliographystyle{apj}


\clearpage

\begin{landscape}
\begin{deluxetable}{lcrrrrr}
  \tabletypesize{\footnotesize}
  \tablecaption{Absorption Line Identifications}
  \tablecolumns{7}
  \tablewidth{0pc}
  \tablehead{\colhead{Observed Wavelength} & \colhead{Identification} &
             \colhead{Rest Wavelength} & \colhead{Redshift} & 
	     \colhead{Rest Frame Equivalent Width} &
             \colhead{Oscillator Strength\tablenotemark{a}} & 
             \colhead{Column Density} \\
             \colhead{(\AA)} & & \colhead{(\AA)} & & \colhead{(\AA)} & &
             \colhead{log (cm$^{-2}$)}
            }
  \startdata
  $7124.23 \pm 0.35$ & \ion{Mg}{2} & 2796.352 & $1.54769 \pm 0.00012$ & 
	$0.18 \pm 0.02$ & 0.612 & $12.63 \pm 0.05$ \\
  $7142.45 \pm 0.42$ & \ion{Mg}{2} & 2803.531 & $1.54766 \pm 0.00015$ &
        $0.08 \pm 0.01$ & 0.305 & $12.58 \pm 0.06$ \\
  7268.49\tablenotemark{b} & \ion{Mg}{1} & 2852.964 & 1.5477 & $< 0.06$ &
    	1.810 & $< 11.7$ 
  \enddata
  \tablenotetext{a}{Ref: \citet{m91}}
  \tablenotetext{b}{The \ion{Mg}{1} upper limits assume a redshift of 
        $z = 1.5477$ and a line width of 20 km s$^{-1}$.}
\label{tab:lineids}
\end{deluxetable}
\clearpage
\end{landscape}

\end{document}